\documentclass[sigconf]{acmart}
\usepackage{booktabs} % For formal tables
\usepackage{enumerate} % Added May 30, 2015 for lists
\usepackage{subcaption}
\usepackage{array}
\usepackage{hyperref}

\newcolumntype{L}[1]{>{\raggedright\let\newline\\\arraybackslash\hspace{0pt}}m{#1}}
\newcolumntype{C}[1]{>{\centering\let\newline\\\arraybackslash\hspace{0pt}}m{#1}}
\newcolumntype{R}[1]{>{\raggedleft\let\newline\\\arraybackslash\hspace{0pt}}m{#1}}

\copyrightyear{2018} 
\acmYear{2018} 
\setcopyright{acmlicensed}
\acmConference[BIRTE '18]{International Workshop on Real-Time Business Intelligence and Analytics}{August 27, 2018}{Rio de Janeiro, Brazil}
\acmBooktitle{International Workshop on Real-Time Business Intelligence and Analytics (BIRTE '18), August 27, 2018, Rio de Janeiro, Brazil}
\acmPrice{15.00}
\acmDOI{10.1145/3242153.3242159}
\acmISBN{978-1-4503-6607-6/18/08}

\begin{document}
\title{Towards Automated Data Integration in Software Analytics}
%\titlenote{Produces the permission block, and
% copyright information}
%\subtitle{Extended Abstract}
%\subtitlenote{The full version of the author's guide is available as
%  \texttt{acmart.pdf} document}

\author{Silverio Mart\'inez-Fern\'andez}
%\authornote{Dr.~Trovato insisted his name be first.}
\affiliation{%
  \institution{Fraunhofer IESE}
  %\streetaddress{P.O. Box 1212}
  %\city{Kaiserslautern}
  %\state{Germany}
}
\email{silverio.martinez@iese.fraunhofer.de}

\author{Petar Jovanovic}
%\authornote{Dr.~Trovato insisted his name be first.}
\affiliation{%
  \institution{Universitat Polit\`ecnica de Catalunya, BarcelonaTech}
  %\streetaddress{P.O. Box 1212}
  %\city{Barcelona}
  %\state{Spain}
}
\email{petar@essi.upc.edu}

\author{Xavier Franch}
%\authornote{Dr.~Trovato insisted his name be first.}
\affiliation{%
  \institution{Universitat Polit\`ecnica de Catalunya, BarcelonaTech}
  %\streetaddress{P.O. Box 1212}
  %\city{Barcelona}
  %\state{Spain}
}
\email{franch@essi.upc.edu}

\author{Andreas Jedlitschka}
%\authornote{Dr.~Trovato insisted his name be first.}
\affiliation{%
  \institution{Fraunhofer IESE}
  %\streetaddress{P.O. Box 1212}
  %\city{Kaiserslautern}
  %\state{Germany}
}
\email{andreas.jedlitschka@iese.fraunhofer.de}

\begin{abstract}
Software organizations want to be able to base their decisions on the latest set of available data and the real-time analytics derived from them. In order to support ``real-time enterprise'' for software organizations and provide information transparency for diverse stakeholders, we integrate heterogeneous data sources about software analytics, such as static code analysis, testing results, issue tracking systems, network monitoring systems, etc. To deal with the heterogeneity of the underlying data sources, we follow an ontology-based data integration approach in this paper and define an ontology that captures the semantics of relevant data for software analytics. Furthermore, we focus on the integration of such data sources by proposing two approaches: a static and a dynamic one. We first discuss the current static approach with a predefined set of analytic views representing software quality factors and further envision how this process could be automated in order to dynamically build custom user analysis using a semi-automatic platform for managing the lifecycle of analytics infrastructures.
\end{abstract}

% Please copy and paste the code. 
%

% \ccsdesc[500]{Information systems~Online analytical processing engines}

% We no longer use \terms command
% \terms{Theory}


\maketitle

\section{Introduction}

Nowadays, the huge amount of data available in companies has increased their interest in applying the concept of "real-time enterprise"\footnote{https://www.gartner.com/technology/research/data-literacy/} by using up-to-date information and acting on events as they happen. In this paper, we envision automated support for the real-time enterprise concept for software organizations by means of applying recent approaches to facilitate data integration tasks.

Currently, software organizations want to be able to base their decisions on the latest set of available data and the real-time analytics derived from them. The software development process produces various types of data such as source code, bug reports, check-in histories, and test cases \cite{Zhang2013}. The data sets not only include data from the development (e.g., GitHub with over 14 million projects), but also millions of data points produced per second about the usage of software (e.g., Facebook or eBay ecosystems). All this data can be exploited with ``software analytics'', which is about using data-driven approaches to obtain insightful and actionable information at the right time to help software practitioners with their data-related tasks \cite{Gall2014}. This improves information transparency for diverse stakeholders. Bearing this goal in mind, we integrate these different data sources as a necessary first step in making this data actionable for decision-making. The integration becomes necessary because %the amount of data is too large and
the inherent relationships in the data influencing the overall software quality are not obvious at first sight. %It is not sufficient to simply collect raw data about current quality issues and provide that data directly to stakeholders (e.g., team leaders, project managers, developers), because the amount of data might be too large and the inherent relationships in the data influencing the overall software quality might be very complex and not obvious at first sight.

Despite its key role, the integration of different software analytics data still presents challenges due to the heterogeneity of the data sources. Not only do data come from sources carrying different types of information, but the same information is also stored in heterogeneous formats and tools. Big Data analytics involves the ingestion of real-time operational data into large repositories (e.g., data warehouses or data lakes), followed by the execution of analytics queries to derive insights from the data, with the final goal of performing business actions or raising alerts \cite{Chandramouli2015}. In a recent systematic review, data integration and final data aggregation were reported as part of the remaining challenges in Big Data analytics \cite{Sivarajah2017}. At the same time, another review in software analytics reported that most of the current approaches are still analyzing only one artifact \cite{Abdellatif2015}, thus not focusing on integrating data from different sources and getting a holistic view.
%In a recent systematic review reporting the challenges in Big Data analytics, data integration and final data aggregation were reported by 12.8\% of their primary studies \cite{Sivarajah2017}. More specifically, another systematic review in software analytics reported that ``most of the [primary] studies'' are still ``analyzing only one artifact'' \cite{Abdellatif2015}, thus not focusing on automatic data integration and aggregation challenges. %Their primary studies analyzing more than one artifact introduce ``direct software statistics like design metrics and change history, simply decorated with some new analytics contributions such as linking team members to the classes they update'' or addressing ``the low-level analytics of source code'' \cite{Abdellatif2015}.
Thus, further research is needed to facilitate the integration of data sources for software analytics driven by the real information needs of end users.

To overcome the heterogeneity of software analytics data coming from different sources, we follow an ontology-based data integration approach in this paper; in particular, we intend to contribute: (a) the definition of an ontology capturing the semantics of relevant data for software analytics; (b) a current static approach for the integration of heterogeneous data sources given a set of predefined analytic views representing software quality factors; and (c) an envisioned approach for the dynamic integration of heterogeneous software analytics data, guided by the specific analytical needs of end users (i.e., information requirements).

The paper is structured as follows: Section \ref{sec:background} presents the related work. Section \ref{sec:ontology} presents an ontology for software analytics. Section \ref{sec:static} presents the implementation of a static approach to implement the integration. Section \ref{sec:dynamic} discusses how the integration could be done dynamically. Finally, Section \ref{sec:conclusions} concludes the paper.
\section{Background and Related Work}\label{sec:background}

\subsection{Software analytics and software quality}

Contrary to the availability of data and its transparency in open source software, tool support for data integration for private companies in commercial systems is just emerging. For instance, we can find some large-scale software companies with their own proprietary development environments, such as Codemine (a proprietary software analytics platform) \cite{Czerwonka2013}, Codebook (a framework for connecting engineers and their work artifacts) \cite{Begel2010} by Microsoft, and Tricorder (a program analysis platform aimed at building a data-driven ecosystem) \cite{Sadowski2015} by Google. Still, these platforms are proprietary and not widely used by others. In addition, companies like Kiuwan, Kovair, and Tasktop have recently started offering software and services for software analytics\footnote{Kiuwan (www.kiuwan.com), Kovair (www.kovair.com), Tasktop (www.tasktop.com)}. Furthermore, very recent research tools are CodeFeedr \cite{Vargas2018} and Q-Rapids \cite{Lopez2018}. Despite these efforts, there are still challenges for companies developing commercial systems to understand how to integrate heterogeneous data sources for software analytics.

Regarding software quality and quality models, a multitude of software-related quality models exist that are used in practice, as well as classification schemes \cite{Klas2009}. One example is the ISO/IEC 25010 standard \cite{ISO}, which determines which quality aspects to take into account when evaluating the properties of a software product. A more recent example is the Quamoco quality model \cite{Wagner2015}, which integrates abstract quality aspects and concrete quality measurements. %Following Quamoco, in \cite{Martinez-Fernandez2018}, abstract quality aspects are also broken down into product factors (attributes of parts of the product that are concrete enough to be measured), assessed metrics (concrete descriptions of how specific product factors should be quantified and interpreted for a specific context), and raw data (i.e. the data as it comes from the different data sources, without any modifications). Therefore, the more abstract levels of quality models depend on metrics coming from diverse data sources. As a consequence, integrating them becomes a necessity.
Nowadays, having operationalized quality models offering actionable analytics from different data (system, process, and usage) is still a challenge.

\subsection{Automating data integration tasks}
Data integration is a well-studied area aiming at facilitating transparent access to a variety of heterogeneous data sources for the end user \cite{Lenzerini02:DataIntegration}. To automate data integration tasks, the use of Semantic Web technologies has been proposed \cite{AbelloRPLNCS15:SemanticWebOLAP}. In particular, we are interested in the use of an ontology for capturing the semantics of heterogeneous data sources due its machine-readable format and the inference capabilities it provides \cite{CalvaneseGHR15:OBDA}. 
The automatic creation of data integration flows has been studied from two main perspectives: 1) starting by analyzing the available data sources (i.e., supply-driven; e.g., \cite{JensenHP04:MDFromRD}),  and 2) starting from the analytical needs of end users and further mapping them to the available data sources (i.e., demand-driven; e.g., \cite{GiorginiRG05:GoalDW}). Others also deploy a hybrid approach combining the previous two ideas \cite{DayalCSW09:DIF4BI,0001RSACN15:Quarry}. 

In this paper, we also envision a hybrid approach for integrating data sources related to software analytics, taking into account the real analytical needs of end users, which may vary over time.

\section{Integrating Data Sources for Software Analytics}\label{sec:ontology}

In this section, we present our software analytics use case that we will use throughout this paper. %First, we introduce an ontology that captures the semantics of relevant data for software analytics, and we show the relationship among its classes for integration. Furthermore, we introduce two examples of information requirements reflecting the real needs of the end user (e.g., project manager).

\subsection{An ontology for software analytics}

\begin{figure}[!b]
	\begin{center}
		\includegraphics[width=0.5\textwidth]{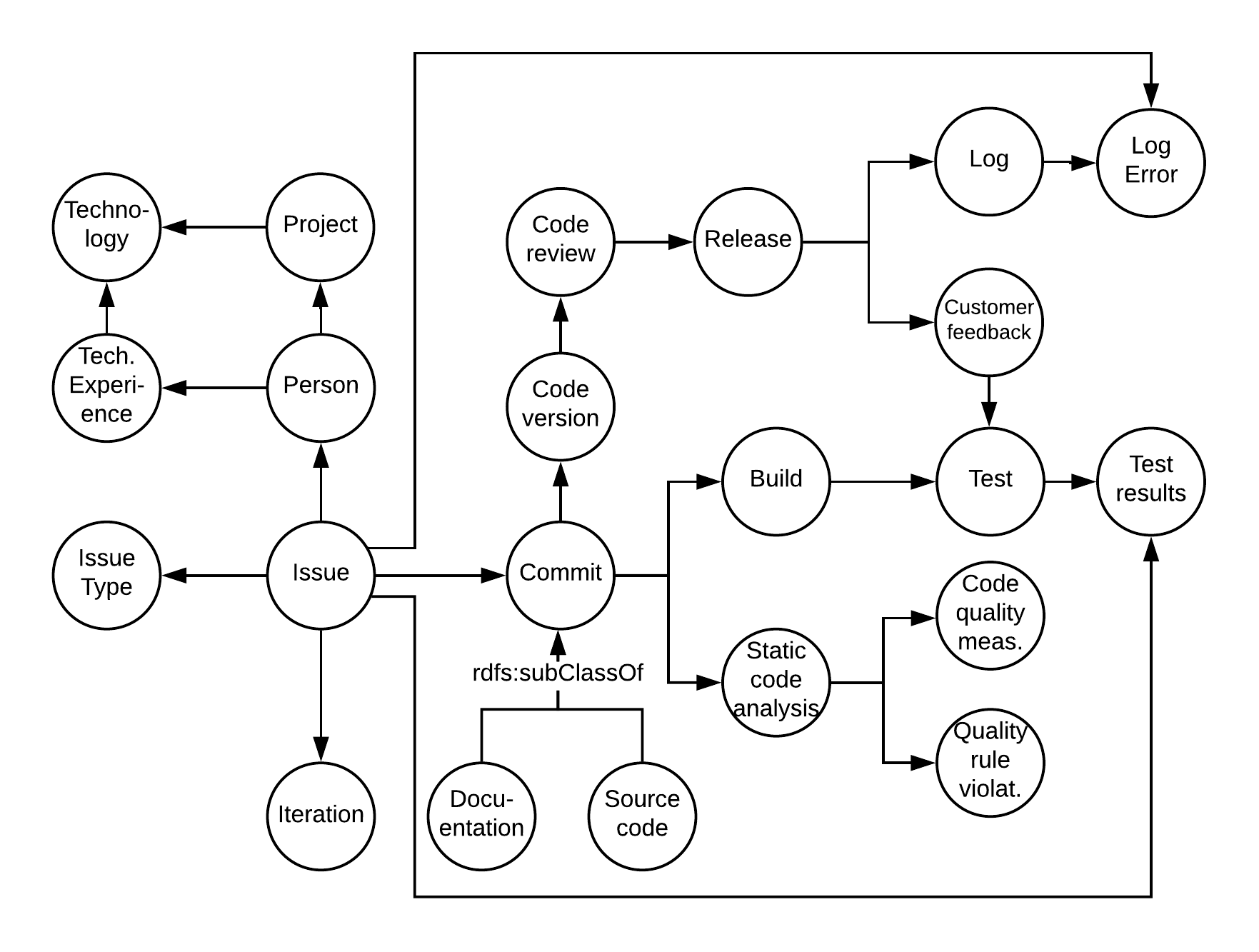}
	\end{center}
	\caption{An ontology capturing the semantics of relevant data for software analytics}
	\label{fig:ontology}
\end{figure}

We introduce an ontology that captures the semantics of relevant data for software analytics (see Fig. \ref{fig:ontology}). In the ontology, each class represents an entity of the software analytics domain. For instance, the class \textit{Issue} represents the issues from issue tracking systems. The ontology is abstract in order to enable generalization and applicability in different software projects. Therefore, the technologies used could differ among projects, whereas the concepts are present in most software development projects. For instance, for issue tracking systems, different companies may use different tools (e.g., Redmine, Jira, or GitLab), but all of them use issue tracking systems as a software development practice. Note that several approaches also propose automated generation of a domain ontology from the desired data sources to support data integration tasks (e.g., \cite{ToumaR015:ontologyConstruction}).

The classes of the ontology in Fig. \ref{fig:ontology} represent data coming from the system either at development time or at runtime. For the sake of simplicity, we omit further ontology details (e.g., datatype properties) in Fig. \ref{fig:ontology}, but explain the main process of how the ontology is built. During software development, we find data about the project and the development, which can be mapped, respectively, to the topics of improving software development process productivity and software system quality presented by Zhang et al. \cite{Zhang2013}. At runtime, we find data about the system behavior and its usage, which can be mapped to the topic of improving the software users' experience. 

%Project data is on tools for managing the backlog containing issues to be done at each sprint as well as their estimated and real effort. Other data is the employees portfolio containing their experience.
First of all, the project owner assigns \textit{Persons} to a certain \textit{Project}, by assessing their \textit{Tech. Experience} with \textit{Technologies}. When the project starts, the \textit{Issues} are defined in an issue tracking system. During sprint planning, the product owner indicates the issues to be resolved in the current \textit{Iteration} (a.k.a. sprint), and estimates the effort with story points. Therefore, it is known to which iteration(s) an issue is assigned. After sprint grooming and planning, the team works on issues, sending \textit{Commits} (a.k.a. changes).

%Development data can be found in repositories (e.g., SVN, git, GitLab), and documentation (e.g., diagrammatic models, Excel files, Wiki pages, Confluence). Other development data relates to code quality (e.g., SonarQube, Klocwork, GitLab), testing and continuous integration (e.g., Jenkins, SpiraTest, QFS).
\textit{Commits} include \textit{Documentation} (such as how-to documents or architectural descriptions) and/or \textit{Source Code}. After a commit is performed, automatic module and system \textit{Builds} containing \textit{Tests} are triggered. Therefore, the \textit{Tests Results} can be associated with a \textit{Commit}. Also, after a commit, \textit{Static Code Analysis} is automatically triggered, producing \textit{Code Quality Measures} (e.g., cyclomatic complexity, lines of code) and checking \textit{Quality Rule Violations}. \textit{Code Reviews} are done over a \textit{Code Version} (a.k.a. branch) instead of intermediate results (i.e., single commits), requiring the results of automatic build and tests, as well as code quality measures and quality rules violations from static code analysis. This way, we know under which conditions a code review is accepted or rejected (e.g., thresholds for code quality measures or test results). If the commit is accepted, it is moved to the main line of a source \textit{Code Version}. The approved source code is heavily tested in nightly builds (including stress testing and stability testing).

%System behavior data may be collected from logs and monitoring tools (e.g., Kibana, elastic, Naggios). Besides, the usage of the software generates useful information such as statistics about how it is used, and feedback directly provided by the end user (e.g., hotlines).
After successful nightly builds, the product owner may decide to create a new \textit{Release}. When the new release is in use, runtime data becomes necessary: \textit{Customer feedback} (normally relating to bugs and system functionalities) and \textit{Logs} (from both end users and executed tests).

Besides the classes, the ontology contains meaningful associations among its classes (see arrows in Fig. \ref{fig:ontology}). For instance, \textit{Commits} can be integrated with \textit{Issues} when the description of the \textit{Commits} includes the ``issue id'' of the related issues.

\subsection{Information requirements for end users} \label{sec:information-requirements}

Analyzing information from single classes is not sufficient for reasoning about quality aspects of the software system, process, or usage. Previous research shows that relevant quality aspects (e.g., maintainability and reliability) contain metrics from heterogeneous data sources, i.e., from several classes of the ontology \cite{Martinez-Fernandez2018}. For this reason, we propose information requirements that consider several classes and help to reason over multiple metrics impacting a quality aspect. Below, we give two examples of information requirements.

\textbf{Information requirement 1} (\textit{IR1}). \textit{Analyze the last release of the software product, per module, ordered by changes, quality rule violations, code quality measures (e.g., complexity, comments, and duplications).}
The goal of IR1 is to improve the code quality of the modules in the next release. The different modules are studied with respect to changes, violations, and code metrics as measurements. Examples of action points to be taken by product owners are: allocating time in the next release to reduce the quality rule violations in a module, or deciding to refactor a highly changed module to make it more stable.

\textbf{Information requirement 2} (\textit{IR2}). \textit{Examine the reliability of a release of the software product in terms of bugs found during testing and errors occurring at runtime, ordered by their resolution time.}
The goal of IR2 is to improve the bug detection of tests. Examples of action points are: improving the test coverage of a unit test, creating further unit tests for a buggy module, testing the software in different contexts, and canceling a release with more bugs than the previous one.

The next two sections report two approaches for data integration using IR1 and IR2 as examples, respectively.

\section{Software Analytics: current static approach }\label{sec:static}

This section explains an implementation for real-time data integration done in the Q-Rapids tool \cite{Lopez2018}. The data flow can be summarized in three steps.

First, during data ingestion, data is gathered during development and at runtime from different data sources. This raw data is ingested into the data stores modeled on the basis of the ontology depicted in Fig. \ref{fig:ontology} and implemented as Elasticsearch indexes.
%, which act as target data stores.
As an example, for IR1, data is gathered from static analysis tools and version control systems and mapped to the following classes: \textit{Code quality measure, Quality rule violation, Commit, Release}.

The high velocity at which data is coming into the system  
%for some classes of the ontology  PETAR: ontology is just a conceptualization no need to mention it here. 
requires the use of Big Data technologies\footnote{Apache Kafka for ingesting (https://kafka.apache.org/), Elasticsearch and Kibana of the Elastic stack for storing and visualizing (https://www.elastic.co/products)} for ingestion and analysis \cite{Lopez2018}. For this reason, the data from each source is ingested with a customized Apache Kafka connector, where the data source is the producer and the connector is the consumer. Then the data is pushed to an index in Elasticsearch, which represents the class of the ontology. For instance, for \textit{Quality rule violation}, there may be multiple connectors for the heterogeneous tools (e.g., SonarQube, CodeSonar, Coverity). The class \textit{Quality rule violation} can be integrated with others such as \textit{Person} (author of the line of code violating a rule), \textit{Code version} (component and line of code fields), and \textit{Commit} (date field), among others.

Second, during data integration and aggregation, two activities are performed to enable the subsequent generation of alerts. In the first activity, quality metrics are computed from the ingested data and further interpreted with a value from 0 to 1. 
%The output of the interpretation of a basic metric is a value going from 0 to 1, being 0 the worst value and 1 the best value regarding quality. 
For instance, the assessed metric `\textit{`Fulfillment of critical/blocker quality rules}'' gives a percentage of the files in the source code without critical or blocker quality rule violations (see \cite{Martinez-Fernandez2018} for details). When the necessary assessed metrics from several data sources are computed, they are stored in another data store as different Elastic indexes. Then the second activity starts. The assessed metrics are aggregated into product factors according to their weight. The weight indicates the relative importance of the assessed metric for the product factor. For instance, for \textit{IR1}, the following assessed metrics are needed: \textit{Non-complex files, Fulfillment of critical/blocker quality rules}, and \textit{Highly changed files}. Following the example, these assessed metrics are aggregated into the \textit{Code quality} and \textit{Blocking code} product factors, as defined in the Q-Rapids quality model \cite{Martinez-Fernandez2018}.

For the implementation, the predefined assessed metrics are translated into Elastic queries, including the formula as well as the execution frequency. Then the query pipeline is executed to compute the assessed metrics and product factors.

Third, product factor alerts can be raised for several reasons, such as a 'bad' value or prediction of a (normalized) product factor. Therefore, alerts act as traffic lights for product factors, only redirecting stakeholders to predefined dashboards when needed. These dashboards contain real-time data to support end users in solving the quality aspect monitored by the product factor. %In the example, the Blocking code product factor has been decreasing in the two last two sprints and an alert is raised. The visualization includes which blocker quality rule violations of type code smell has to be solved. By selecting them, the user sees the module, line of code, and a explanation of the code smell problem.

\begin{figure}[!t]
	\begin{center}
		\includegraphics[width=0.5\textwidth]{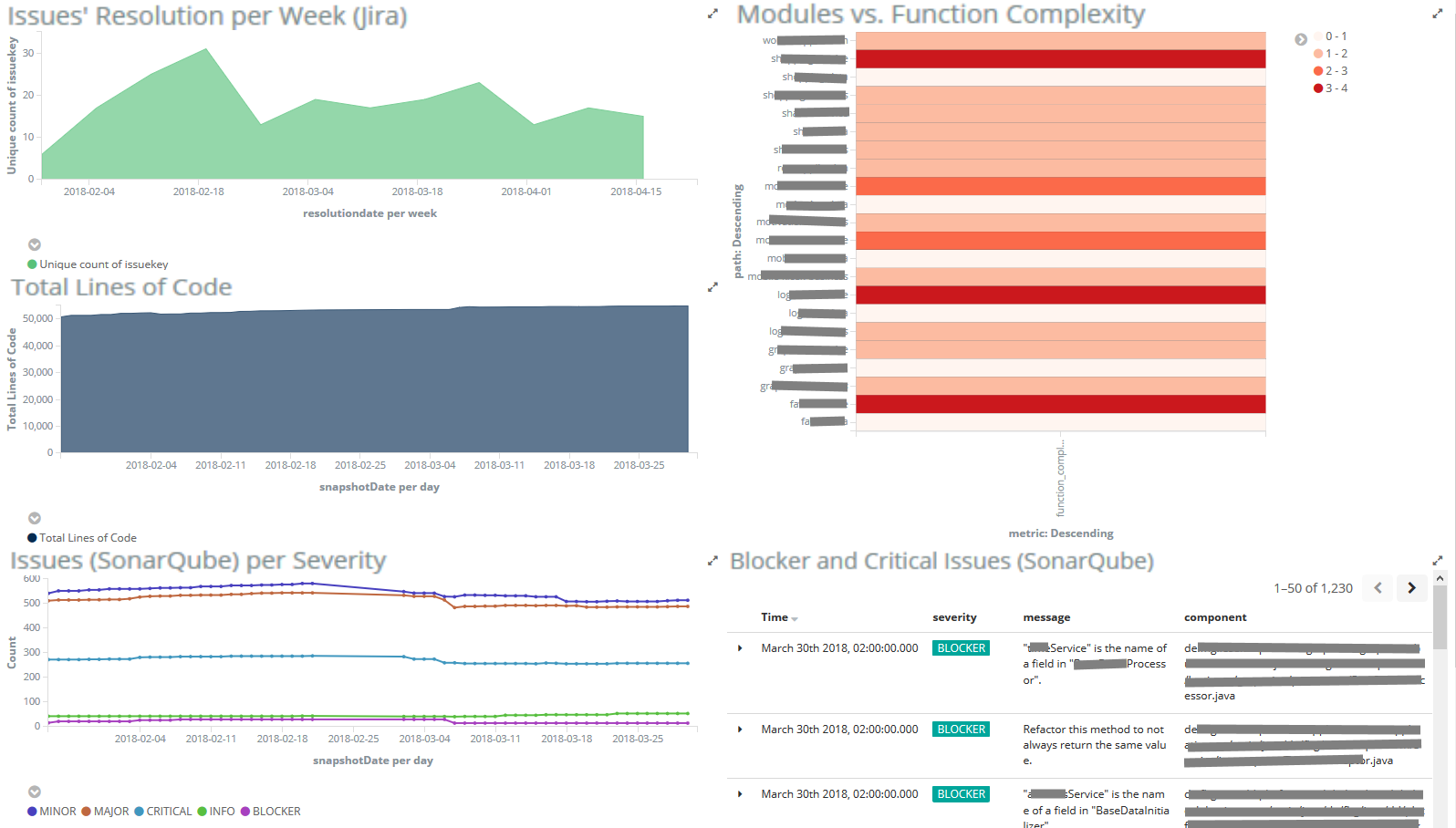}
	\end{center}
	\caption{Example of raw data visualization for IR1}
	\label{fig:kibana}
\end{figure}

These dashboards are implemented as Kibana objects (following Kibana terminology, they are dashboards consisting of visualizations and searches). 
As an example of IR1, an alert for \textit{blocking code} can be raised because this product factor has been decreasing in the last sprints. Then the user is redirected to the dashboard shown in Fig. \ref{fig:kibana}, with the following data about the selected release:
\begin{itemize}
\item ``Modules vs. Function Complexity'': It shows a heat map with the average function complexity of the modules. Users can drill down and see a list of files that should be checked.
\item ``Issue Resolution per Week'' and ``Total Lines of Code'': By showing the velocity of the development team based on the resolution of issues and the evolution of the size of the code, the users can see if they are correlated with \textit{blocking code} problems.
\item ``Issues per Severity'' and ``Blocker Issues'': This shows the quality rule violations according to severity. In addition, a list of blocker and critical violations is shown that we suggest should be taken care of. Within this list, the user can filter the violations (e.g., rules of the type ``code smell''). Also, by opening them, the user sees the module, line of code, and an explanation of the code smell problem. Therefore, after an alert has been raised, the user can see the details of each violation in order to take actions in real time.
\end{itemize}

\section{Towards Automated Software Analytics}\label{sec:dynamic}

In the previous section, we saw how the integration of data from multiple sources can be beneficial for extracting actionable analytics for the software process, system, and usage. However, as can be seen, such integration requires considerable manual effort on the part of the designer to
%, who starting from a given information requirement, needs to identify the data sources that can be used for answering the requirement, implement the data flows for extracting and integrating such data, model the target data stores, and 
integrate, transform, and prepare the data to be plugged into the desired analytics or visualization tool for further exploitation. Furthermore, given that information requirements may often be ambiguous and/or incomplete, the above process may undergo several rounds of reconciliation and redesign until the real information needs of an end user are finally met.

Considering that the analytical needs of different stakeholders, such as team leaders, project managers, or developers, are different and can, moreover, change dynamically over time, the proposed manual process may become an overburdening bottleneck. 

Therefore, we envision a system that would apply and extend existing approaches in order to automate the process of building data integration flows for the field of software analytics. In particular, one such system, called Quarry \cite{0001RSACN15:Quarry}, provides an automated, end-to-end solution for assisting the users of various technical skills in managing the design and deployment of analytical infrastructures, i.e., target data stores and data-intensive flows like extract-transform-load (ETL) processes. Quarry starts from high-level information requirements (e.g., IR1 and IR2 in Section \ref{sec:information-requirements}) and semi-automatically
%provided by end users through the graphical assistance tool. 
%Quarry then automatically validates the mapping of the provided information requirement over the available data sources, explores the ontology to find how the required concepts are related, and finally, based on the gathered information, 
generates target data stores and a data flow to integrate data from different data sources and prepare them for further exploitation. 

For instance, a user interested in examining the reliability of a release in terms of issues of the type bugs found during testing and errors occurring at runtime detected in logs (see IR2 in Section \ref{sec:information-requirements}) would pose such a requirement by selecting the ontological concepts in the graphical tool and adding additional query information.

\begin{figure}[h!]
	\vspace{-10pt}
    \center
    \includegraphics[width=1\linewidth]{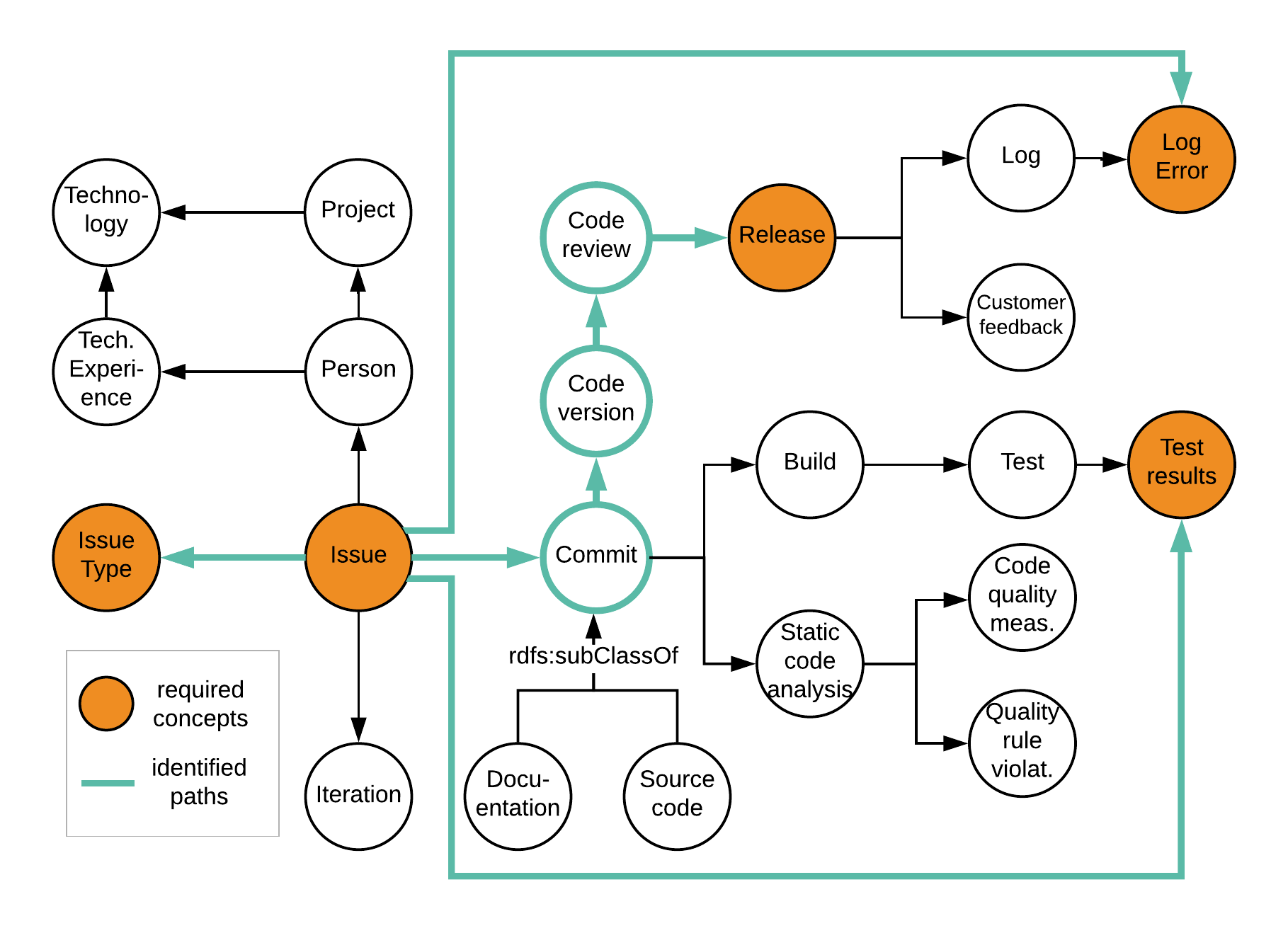}
    \vspace{-10pt}
    \caption{Identified ontological concepts (i.e., classes and properties) for IR2}    
    \label{fig:ontology-paths}
    
\end{figure}

\begin{figure}[h!]
    \center
    \includegraphics[width=1\linewidth]{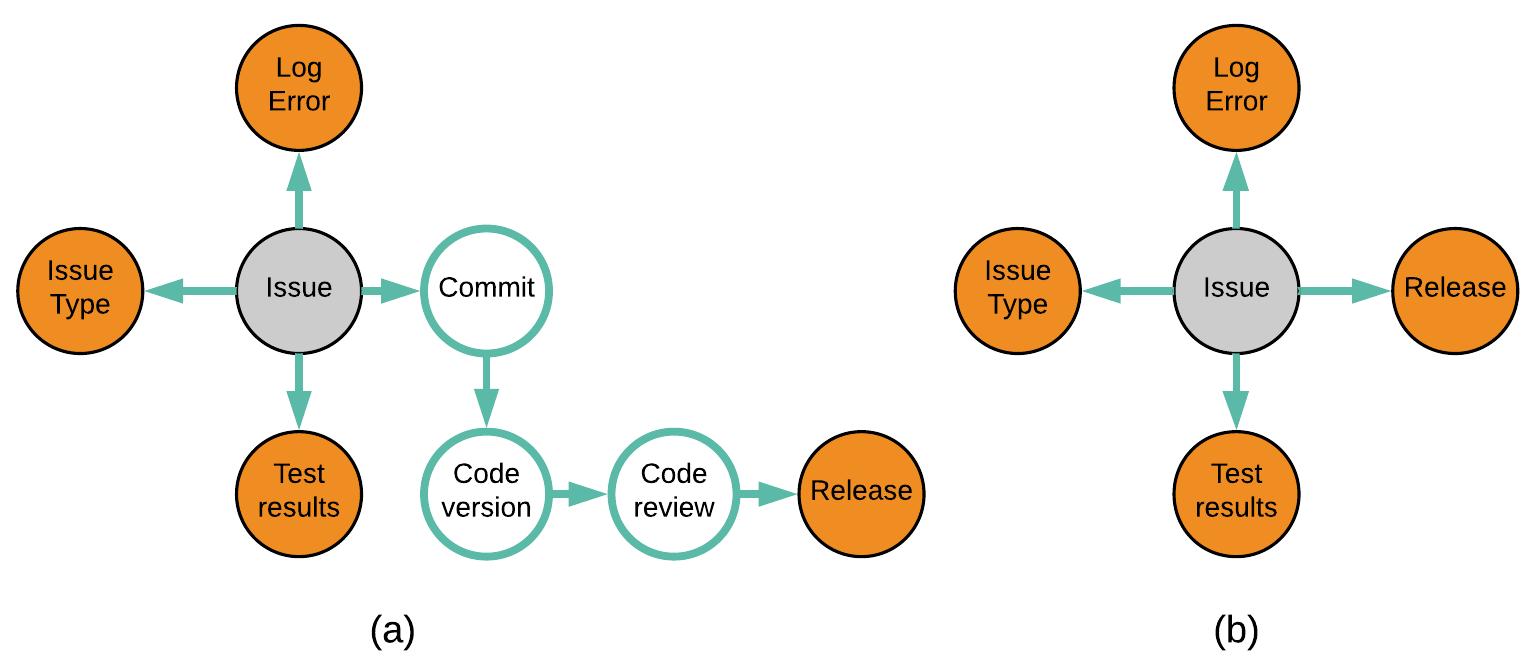}
    \vspace{-20pt}
    \caption{(a) Extracted ontology subset,  and (b) generated schema of the target data store}\label{fig:extracted-ont-schema}    
\end{figure}

Looking at the ontology in Fig. \ref{fig:ontology}, the user can easily express such a requirement in his/her own vocabulary using Quarry's graphical tool, and hence Quarry will identify the ontological concepts requested by the user (i.e., \textit{Release}, \textit{Issue}, \textit{Issue Type}, \textit{Test results}, \textit{Log Errors}; see Fig. \ref{fig:ontology-paths}). Moreover, the user can express restrictions regarding the given concepts (e.g., issue being of the type ``bug'') and aggregations (e.g., count the issues). Starting from the identified concepts, Quarry explores the ontology and finds the paths through which such concepts can be related. Notice that in order to satisfy the summarizability properties \cite{LenzS97:SummarizabilityOLAP}, which are needed to correctly answer the posed requirement, these paths must respect multidimensional integrity constraints (i.e., have ``to-one'' cardinalities). Thus, the selected paths are shown in Fig. \ref{fig:ontology-paths}. Going from here, Quarry extracts the subset of ontology concepts needed to answer the given requirement (see Fig. \ref{fig:extracted-ont-schema}(a)), and generates the schema for the target data store (Fig. \ref{fig:extracted-ont-schema}(b)).

In the background, Quarry also generates a complete data flow at the same time, which (1) extracts data from different data sources to which the identified ontology concepts map; (2) integrates these data following the identified paths in the ontology (by means of performing joins among the mapped data); and, finally, (3) applies restrictions (e.g., \textit{Issue Type} = ``\textit{bug}'') and aggregations (e.g., \textit{count}(issues)) that the user may have expressed through his/her requirement. 

To make all this work, Quarry also provides a deployment module, which can be extended to translate the generated constructs into the desired formats ready for deployment. On the one hand, the target data store schema can be translated either into a set of standard database tables implementing relational OLAP or into a set of Elasticsearch indexes as seen in the previous section. On the other hand, a data flow can be represented either as an ETL process implemented as a set of SQL views or in a proprietary format of an ETL tool (e.g., Pentaho Data Integration PDI), or as a query for immediately retrieving the data (e.g., SQL or Elasticsearch).

Finally, by deploying the generated analytical infrastructure (i.e., target schema and corresponding data flow), the system is ready to integrate and transform the input data coming from different data sources and to store it following the schema model in order to satisfy the user's information needs and prepare the data for further exploitation (e.g., real-time data visualization; see Figure \ref{fig:kibana}).

\section{Conclusions and Future Work}\label{sec:conclusions}

The automatic integration of data from different sources is still a challenge for the software analytics domain. In this position paper, we defined an ontology capturing the semantics of software analytics data sources. Furthermore, we showed the current static approach to data integration in the Q-Rapids tool \cite{Lopez2018}. Finally, we envisioned a dynamic approach for the generation of dashboards with actionable analytics defined by the end users.

In the dynamic approach, the end users could, based on their own analytic needs, easily build data integration flows in order to prepare software analytics data to be plugged to external analytical tools. This will enable end users to explore and understand holistic software quality aspects by transparently considering different sources of information.

An immediate future direction is to conduct a detailed case study on applying automated approaches like Quarry \cite{0001RSACN15:Quarry} to the software analytics use case, with the aim of validating the benefits envisioned in this position paper. 

%Using Quarry enables the dynamic generation of adaptive dashboards that support data exploration and infrastructure. 
%he aim of such case study will be to validate the benefits envisioned in this position paper. 

% actionability is one of the key promises of software analytics over simple repository mining (which is more descriptive)
% Everything in the same place (instead of distributed data sources) improves transparency
% Future work: Implementation of Enhancements using Quarry for more dynamic generation of adaptive dashboards that support data exploration and infrastructure
% The diverse frequency of ingestion in the data sources.

\begin{acks}
We thank Axel Wickenkamp for the implementation of the Q-Rapids tool and Sonnhild Namingha for proofreading the paper. This work was supported by the Q-Rapids project (H2020, No. 732253) and the ERCIM Fellowship Programme.
\end{acks}

\bibliographystyle{ACM-Reference-Format}
%\bibliography{acmart}
\bibliography{birte2018}

\end{document}